\documentstyle[aps,epsf,epsfig,psfig,float,latexsym]{revtex}  

\newcommand{\bi}{\bibitem}  
  
\newcommand{\be}{\begin{equation}}  
\newcommand{\ee}{\end{equation}}  
\newcommand{\ba}{\begin{eqnarray}}  
\newcommand{\ea}{\end{eqnarray}}

\newcommand{\beq}{\begin{equation}}  
\newcommand{\eeq}{\end{equation}}  
\newcommand{\beqa}{\begin{eqnarray}}  
\newcommand{\eeqa}{\end{eqnarray}}

\begin{document}  
\title{  
\begin{flushright}  
{\normalsize {FZJ-IKP(TH)-2000-12}}\\  
{\normalsize{}}  
\end{flushright}  
\vspace{2 cm}  
$J/\Psi \to \phi \pi\pi (K\overline{K})$ decays, chiral dynamics 
and OZI violation  
}  
\vspace{2.5 cm}  
\author{Ulf-G. Mei{\ss}ner\footnote{email: Ulf-G.Meissner@fz-juelich.de},  
J. A. Oller\footnote{email: j.a.oller@fz-juelich.de} \\ [1cm]  
{\small {Forschungszentrum J{\" u}lich, Institut f{\" u}r Kernphysik  
(Th), D-52425 J{\" u}lich, Germany}}}   
\maketitle  
  
\thispagestyle{empty}  
  
\vspace{2cm}  
  
\begin{abstract}  
We have studied  the invariant mass distributions of the $\pi\pi$ and 
$K \overline{K}$ systems for invariant masses up to 1.2 GeV from the $J/\Psi \to \phi \, 
\pi\pi(K\overline{K})$ decays. The approach exploits the connection between these 
processes and the $\pi\pi$ and $K\overline{K}$ strange and non-strange scalar form 
factors 
by considering the $\phi$ meson as a spectator.
The calculated scalar form factors are then matched 
with the ones from next-to-leading order chiral perturbation theory,
including the calculation of the the $K\overline{K}$ scalar form factors.
Final state interactions in the $J/\Psi \rightarrow \phi \pi\pi(K\overline{K})$ processes 
are taken into account as rescattering effects in the system of the two
pseudoscalar mesons. A very good agreement 
with the experimental data from DM2 and MARK-III is achieved. Furthermore, making use of
SU(3) symmetry, the S-wave 
contribution to the $\pi^+\pi^-$ event distribution in the
$J/\Psi \rightarrow \omega \pi^+\pi^-$ reaction is also given and the data up 
to energies of about 0.7 GeV are reproduced. 
These decays of the $J/\Psi$ to a vector and a pair of 
pseudoscalars turn out to be very sensitive to OZI violating physics which we parameterize
in terms of a direct OZI violation parameter and the chiral perturbation
theory low energy constants $L_4^r$ and $L_6^r$. 
These constants all come out very different from zero,
lending further credit to the statement that the OZI rule is subjected to large
corrections in the scalar $0^{++}$ channel. 
\vspace{1cm}  
  
  
\end{abstract}  
\newpage  
  
\section{Introduction}  

\begin{figure}[htb]
\centerline{
\epsfysize=1.7 in 
\epsffile{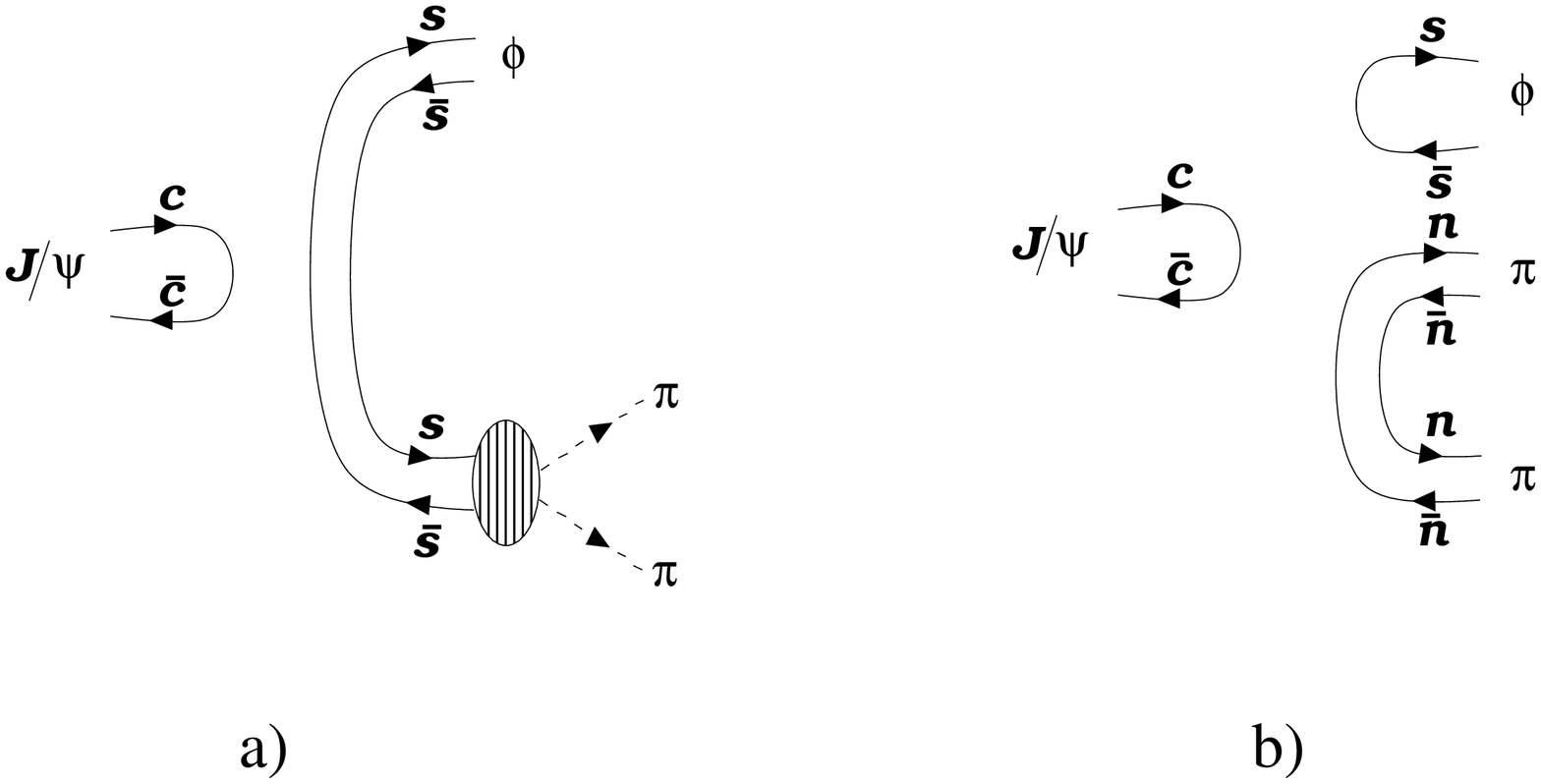}
}
\vskip 1cm 

\caption[]{Quark line diagrams for decay $J/\Psi$ into the $\phi$ and a meson
pair ($\pi\pi$ or $K\overline{K}$). The quark flavors are explicitly given,
$n$ refers to the light non--strange $u,d$ quarks. The hatched blob in a)
depicts the final state interactions in the coupled $\pi\pi/K\overline{K}$ system.
\label{fig:OZI}}
\end{figure}

The decays of the $J/\Psi$ into a $\phi$ meson and Goldstone boson
pair ($\pi\pi$ or $K\overline{K}$) can be used to investigate the 
dynamics of the interacting pseudoscalars. In particular, it was
argued in Ref.\cite{pen} that these data together with data from
pion--pion scattering (and from others sources) force the $f_0(980)$ to have 
a pole structure different to the one required by a $K\overline{K}$ molecule \cite{isgur2}. 
This interpretation has been challenged, e.g. in
the J\"ulich meson--exchange model where the $f_0 (980)$ emerges~\cite{Juel} as a 
$K\overline{K}$ bound state. Furthermore, in this reference the authors are also able to 
reproduce the data associated with the previous 
$J/\Psi$ decays within the same formalism than the one employed in Ref. \cite{pen}, but
making use of their own strong amplitudes. On the other hand, as will be the topic of 
this investigation, these data can be used to study the violation of the
Okubo--Zweig--Iizuka (OZI) rule in the scalar ($0^{++}$) channel. This rule
is only well founded in the large $N_c$ limit of QCD, with $N_c$ the
number of colors, since OZI violating processes are described by
suppressed non--planar graphs~\cite{witten}. Still, on a purely
phenomenological level this rule works astonishingly well, with the
exception of the scalar channel, as argued e.g. in
Refs.\cite{isgur},\cite{stern},\cite{isgth}. To be more precise, the decay $J/\Psi
\to \phi M \bar M$ (where $M\bar M$ denotes the pseudoscalar meson pair) is OZI
suppressed to leading order, cf. Fig.\ref{fig:OZI}a, but has an additional doubly
OZI suppressed contribution depicted in Fig.\ref{fig:OZI}b. In our approach, both
these pieces are taken into account. In fact, it will turn out that the
second contribution can not be neglected if one wants to achieve an accurate
description of the data. On the other hand, it is mandatory to have a very
precise description of the final state interaction in the coupled
$\pi\pi / K\overline{K}$ system (as indicated by the shaded blob in Fig.\ref{fig:OZI}a)
before one can ask such detailed
questions. As can be seen from Fig.\ref{fig:OZI}, 
the crucial ingredient in the reaction at hand are the
expectation values of the scalar--isoscalar condensates in the pion
and the kaon, i.e the so-called scalar form factors. These can be
calculated at low energies in chiral perturbation theory (CHPT), which
is the effective field theory of the Standard Model. In our case, the
dimeson system can have energies up to 2~GeV and we thus employ
unitarity constraints to get a precise description of these scalar
form factors also at higher energies, demanding furthermore matching to the CHPT expressions
in the low energy domain.
Because of this matching procedure, the large $N_c$ suppressed low
energy constants $L_4^r$ and $L_6^r$ of the next--to--leading order
effective chiral Lagrangian can be determined in the process we are
considering. It has been argued before that so far no direct determinations but
rather large $N_c$ inspired estimates have been done, see
e.g. Refs.\cite{stern},\cite{gl}, with the exception of more
recent work presented in Refs.\cite{stern,mouss,mouss2}. Nevertheless, as we will 
discuss in much more detail below, a rather definite  determination of 
$L_4^r$ can be obtained by considering ${\mathcal{O}}(p^6)$ CHPT results 
\cite{bij,bij98}. 

\medskip \noindent
To be more specific, to address the problem
of the  final state interactions in the coupled  
$\pi\pi$-$K\overline{K}$ system, we make use of the results obtained in   
Ref.\cite{npa}. In this paper it was clearly established that the  
scattering  data of the $0^{++}$ $I=0, \, 1$ ($I$ denotes the total
isospin of the dimeson system)  
sectors up to centre--of-mass energies of 1.2 GeV are a reflection of  
the strong rescattering  effects between the lightest  
pseudoscalars ($\pi\pi$, $K\overline{K}$ for $I=0$ and $\pi\eta$ and
$K\overline{K}$ for $I=1$).  The  
approach was based on  Bethe-Salpeter equations using  the lowest order 
CHPT amplitudes\cite{wein,gl1,gl} as the driving potential. The fact   
that one can generate the resonance states for those channels via loop  
physics, i.e. rescattering, is a clear signal of the large deviations from  OZI rule  
in the $0^{++}$ sector, see also Refs.\cite{stern,isgur}.  Such a mechanism has
been advocated since long, for a pedagogic discussion see Ref.\cite{Ulfcnpp}.
On the other hand, it is well known that there is an on--going  
controversy concerning  the nature of the scalar  
resonances $f_0(980)$ and $a_0(980)$. This controversy originates from  
the observation that   
there are several different models to deal with the $I=0 , \,1$ scalar   
sector, all of them reproducing the scattering data up to some extend, but with 
different  
conclusions with respect to the origin of the underlying dynamics. In particular, in  
Refs.\cite{Torn,van,jaf} these resonances are considered of preexisting origin   
while in Ref. \cite{isgur2} they appear as meson--meson resonances originated by a   
potential. Also in Ref. \cite{amsler} it is advocated for the solution that the  
$a_0(980)$, $f_0(980)$ are exotic resonances, that is, not simply $q\bar{q}$, while the   
preexisting $q\bar{q}$ scalar nonet should be heavier, around 1.4 GeV or so. 
Other interesting approaches to this problem are the relativistic quark model
with an instanton induced interaction of the Bonn group~\cite{Bonn}, 
the J\"ulich meson--exchange approach~\cite{Juel} or the use of QCD sum rules 
\cite{steele}.  With respect to this controversy, the contribution of the work in Ref.
\cite{npa} is very valuable
since, at least,  the infinite series of diagrams there considered should
appear in the whole S--wave partial wave  amplitudes calculated to all orders in CHPT. 
The conclusions of Ref.\cite{npa} where generalized in Ref.\cite{nd}. In that paper, the most  
general structure of a partial wave amplitude when the unphysical cuts are neglected was  
established. In particular, in this paper  explicit s-channel  
resonance exchanges were included together with the lowest order CHPT
contribution and the whole  
$SU(3)$ connected scalar sector with $I=0,1/2,1$ was studied. In particular, it was   
shown that the amplitudes of Ref.\cite{npa} appear as a particular case when removing the  
explicit tree level resonance contributions. It was observed that the  
lightest $0^{++}$ nonet is of dynamical origin, i.e. made up of meson--meson  
resonances, and is formed by the $\sigma(500)$,   
$\kappa$, $a_0(980)$ and a strong contribution to the physical $f_0(980)$. On the other  
hand, the preexisting scalar nonet would be made up by an octet around 1.4 GeV and a  
singlet contributing to the physical $f_0(980)$ resonance.  With respect to this last  
point, as discussed in Ref.\cite{nd}, the inclusion of a preexisting contribution to the  
$f_0(980)$ was considered in order to be able to  
reproduce the data on the inelastic $\pi\pi\to K\overline{K}$ cross section\footnote{In the last  
edition of the PDG tables~\cite{pdg} it is argued that, possibly, the previous experiments have  
a much larger uncertainty than previously given in the corresponding publications.} when   
including also the $\eta\eta$ channel. However, if this channel is not considered, one can   
reproduce the strong scattering data, including also the previous experiments on the  
inelastic $\pi\pi \rightarrow K\overline{K}$ cross section, without including such preexisting   
contribution. Finally, in Ref.\cite{nd} the  
contribution in the physical region of the unphysical cut contributions were estimated  
up to $\sqrt{s}\approx 800$ MeV to be just a few per cents making use of the results of   
Ref.\cite{ulfkpi}, which apply below that energy.   
  We will use the formalism of Ref.\cite{npa} whose partial wave amplitudes have been   
also tested in many other reactions.  As pointed out in  
Ref.\cite{pen}, to obtain a consistent picture of the scalar sector,  
one also has to study other reactions in which the $0^{++}$   
amplitudes have a possible large influence via final state  
interactions. In this way one can complement the deficient  
information coming from the direct strong S--wave  scattering data and distinguish between  
available models. In Ref. \cite{gamma} all the whole set of photon  
fusion reactions $\gamma \gamma \rightarrow \pi^0\pi^0$, $\pi^+ \pi^-$, $K^+ K^-$, $K^0   
\overline{K}^0$ and $\pi^0 \eta$ were reproduced in an unified way from threshold up to   
$\sqrt{s}\approx 1.4$   
GeV making use of Ref.\cite{npa} to take care of the final state interactions. The free  
parameters present in Ref.\cite{gamma} were fixed by their values from the PDG  
\cite{pdg}. In Ref.\cite{marco}, making use of the formalism set up in Ref. \cite{plb}  
to study the still unmeasured $\phi \to \gamma K^0 \overline{K}^0$ decay,   
the reactions $\phi \to \gamma \pi^0 \pi^0$, $\gamma \pi^+ \pi^-$ and $\gamma  
\pi^0 \eta$ were predicted. These predictions were nicely confirmed, almost  
simultaneously, by a recent experiment in Novosibirsk \cite{novo}.   
In this manuscript, we will consider yet another applications of the strong  
amplitudes  calculated in Ref.\cite{npa} by studying the $J/\Psi \to   
\pi\pi(K\overline{K})$ decays. In this way, the present study together with the whole set   
of works\cite{npa,gamma,plb,marco} offer  a unique   
theoretical approach to the scalar sector able to discuss all these reactions in an unified   
way. This is achieved without including new elements ad hoc for each  
reaction, because all these processes are related by the use of an  
effective theory description  that combines CHPT and unitarity constraints.  

\medskip\noindent  
Our manuscript is organized as follows. In section~\ref{sec:model} we
develop and justify
a simple phenomenological model for the transition of the $J/\Psi$ into the $\phi$ meson 
and a pseudoscalar meson pair. This model is generalized in section \ref{sec:res} by 
making use of flavor SU(3) symmetry and then applied to the $\omega$ case. Section 
\ref{sec:pikff} is devoted to the calculation of the
scalar pion and kaon form factors for the non--strange as well as the strange scalar
density to one loop. Then we employ the unitarization procedure discussed
before to obtain a description of these quantities up to energies of 1.2~GeV.
We also perform the matching of these expressions to the one loop CHPT ones
to be consistent with the constraints of the broken chiral symmetry of QCD at
energies below approximately 0.5~GeV. The results are presented and discussed
in section~\ref{sec:res} where we also generalize section \ref{sec:model}. Our conclusions and outlook are given in section~\ref{sec:concl}.

\section{Modeling $J/\Psi \to \phi\pi\pi\;, \phi K\overline{K}$ decays}  
\label{sec:model}  
We will calculate the S-wave contribution to the invariant mass distributions of the  
$\pi\pi$ and $K\overline{K}$ systems in the $J/\Psi \to \phi \pi\pi(K\overline{K})$ decays. Taking care of the  
final state interactions of three particles can be simplified to a large extend if only  
two of the final state particles undergo strong interactions, and the third is merely a  
spectator. We will assume that this is the case in our present problem and we will take  
the $\phi$ as the spectator. This is certainly very well sounded for the $\phi \pi\pi$  
system since the $\phi \pi$ interaction is very weak as required by the OZI~\cite{OZI}   
rule. On the  
other hand, the situation is not so clear with respect the $\phi$ and the kaons.   
Nevertheless, at the energies in which the kaons become important, above the $K\overline{K}$  
threshold, the experimental mass distribution is completely dominated by   
the $f_0(980)$ resonance and this state is a two body effect emerging  
from the coupled $\pi\pi$ and $K\overline{K}$ systems, as discussed  
already in the introduction. Since we are only considering a small  
range of energies above the $K\overline{K}$ threshold, this  
approximation should be justified.  
  
 \medskip\noindent  
We therefore describe the  transition from the $J/\Psi$ to the  
$\phi$+2 Goldstone bosons system by an effective Lagrangian based on the following 
phenomenological arguments: 1) The already discussed spectator role of the $\phi$ 
resonance and 2) the $\pi\pi$ and $K\overline{K}$ invariant event
distributions, which will be shown later, 
seem to be clearly dominated by the S-wave 
contribution, although these experimental data have not yet been
subjected to a partial-wave analysis. These 
{\it{experimental}} facts, together with Lorentz invariance, can be easily incorporated 
in the formalism just by writing the interaction vertex of the $J/\Psi$ resonance  
with the $\phi$ meson and some scalar source $S$ with vacuum quantum numbers,  
$J^{PC}=0^{++}$ with spin $J$, parity $P$ and  charge conjugation $C$, as:    
\be  
\label{lagS}  
g\, \Psi_\mu \phi^\mu S  
\ee  
with $g$ a real coupling constant. We briefly discuss why other
possible structures involving derivatives on the various fields
should be suppressed. The $J/\Psi$ is very heavy and thus can be
considered a static source. Derivatives acting on the $\phi$ and the
scalar source $S$ can be combined to the invariant structure $\Psi^\mu
(\partial_\nu \phi_\mu - \partial_\mu \phi_\nu ) \partial^\nu S$. This
leads to a vertex of the form $\epsilon^\mu_\Psi \epsilon_\mu^\phi
\, p_\phi \cdot (q_1+q_2)$, with $q_{1,2}$ the momenta of the two
Goldstone bosons and $p_\phi$ the momentum of the $\phi$ meson. 
However, due to momentum conservation, we have
$p_\phi \cdot (q_1+q_2) = (M_\Psi^2-M_\phi^2-s)/2$, with  $s=(q_1+q_2)^2$ 
the total two Goldstone boson energy squared. Due to the large value of the $J/\Psi$ mass, this
combination of momenta is essentially constant for the Goldstone boson
energies considered here, $\sqrt{s} \le 1.2\,$GeV. Such terms become more
important at higher di--pion (kaon) energies.
Therefore, we can generically write such
type of higher order corrections to Eq.(\ref{lagS}) in the form
\be
\epsilon^\mu_\Psi \epsilon_\mu^\phi \, f\left( (q_1+q_2)^2, p_\phi \cdot
(q_1+q_2)\right)~,
\ee
where the function $f(\ldots)$ essentially only depends on the first argument. Such terms that
depend on $(q_1+q_2)^2$ can be derived from an interaction of the type
$\Psi^\mu \phi_\mu \partial_\nu \partial^\nu S$. Such structures lead
to a weak $s$--dependence of the constant $g$ and/or the parameter
$\lambda_\phi$ defined below. We have checked that such (weak) energy dependencies do not
change any of the conclusions obtained when treating $g$ and $\lambda_\phi$
as energy independent. Another possible higher order term  of the  
form $\Psi_\mu \phi_\nu \partial^\mu \partial^\nu S$ giving rise to the
vertex $\epsilon^\mu_\Psi \epsilon_\nu^\phi \,(q_1+q_2)_\mu (q_1+q_2)^\nu$.
Such couplings can also have an S--wave contribution, which can be obtained
by properly summing over the pertinent polarization vectors. Again, due to
the large mass of the $J/\Psi$, such terms are only weakly $s$ dependent
and can be treated along the lines outlined before. More complicated
structures can always be brought into some linear combination of the ones
just discussed or have no S--wave component. These considerations not only
show that our ansatz Eq.(\ref{lagS}) is quite sensible in the energy range
considered here but also that corrections to it can be worked out consistently.

\begin{figure}[htb]
\centerline{
\epsfysize=1.2 in 
\epsffile{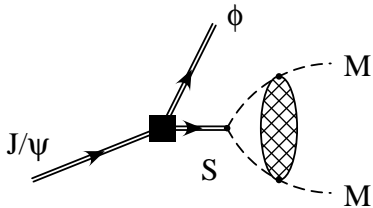}
}
\vskip 1cm 

\caption[]{Anatomy of the $J/\Psi$ decay into a $\phi$ and a Goldstone
  boson pair. $S$ is the interpolating scalar field described in the text.
  The cross--shaded blob symbolizes the final state interactions in the 
  coupled $\pi\pi/K\overline{K}$ system.
\label{fig:anat}}
\end{figure}

 From the OZI rule, which can also  
be seen as a result of the large $N_c$ expansion of QCD~\cite{witten}, and the
experimental absence of any clue indicating a non negligible interaction between the
$\phi$ and the pions, one should expect  
that this scalar source $S$ would be simply made of strange quarks,  
i.e. $S \sim \bar{s}s$. However, it is known that the $\phi$ also  
decays into non--strange mesons and furthermore, there are strong  
arguments to believe  
that large violations of the OZI rule (and of the large $N_c$ limit of  
QCD) are manifest in the $0^{++}$ sector~\cite{stern,isgur,nd,npa}, as
discussed in the introduction. As a result, we  
will consider a more general scalar source $S$, that also has a  
contribution of the  form $\lambda_\phi \bar{n}n$, where  
\be\label{nquarks}  
\bar{n}n = {1 \over \sqrt{2}} \, (\bar{u}u+\bar{d}d)  
\ee   
parameterizes the contribution  
from the non--strange quarks and $\lambda_\phi$ is just a constant measuring the  
relative strength of this contribution with respect to the strangeness  
component $\bar{s}s$. Already at this point we stress that the choice  
$\lambda_\phi \neq 0$ will be justified a posteriori by the results  
presented below. Therefore, we use  
\be  
\label{S}  
S=\bar{s}s+\lambda_\phi \,\bar{n}n  
\ee  
Consequently, it follows from Eqs.(\ref{lagS}) and (\ref{S}) that the transition   
matrix element for the process $J/\Psi \rightarrow \phi \, M\bar{M}$ is given by:  
\be  
\label{tree}  
T = \epsilon(\Psi;\,\rho)_\mu  \epsilon(\phi;\,\rho')^\mu  
\,\langle0|\left(\bar{s}s+\lambda_\phi\,\bar{n}n\right)|M\bar{M}\rangle^*  
\ee   
where $|0\rangle$ is the vacuum state, $\epsilon(\Psi;\,\rho)$  is the polarization   
four--vector of the $J/\Psi$ resonance with polarization $\rho$ and analogously  
$\epsilon(\Psi;\,\rho')$ is the polarization four--vector of the $\phi$  
resonance, and the $^*$ denotes complex conjugation. Note that in this  
equation we are implicitly assuming  that the $\phi$ is a spectator as discussed before.  
 The scalar source $S$  
couples to the two meson system, in which the rescattering (final  
state interactions) appear. The anatomy of our model is depicted in  
Fig.\ref{fig:anat}.  As discussed below, invoking SU(3) symmetry,
we will also apply this approach to
the S--wave contribution of the $J/\Psi \to \omega \pi\pi$ decay to further
constrain the description of the measured event distributions. 
  
\section{Coupled channel pion and kaon scalar form factors}  
\label{sec:pikff}
  
As a consequence of Eq.(\ref{tree}), our problem is reduced to calculate the matrix  
elements $\langle0|\bar{s}s|M\bar{M}\rangle$ and $\langle  
0|\bar{n}n|M\bar{M}\rangle$,   
which correspond to the strange and non-strange 
isospin zero ($I=0$) scalar form factors.  
  
\subsection{Definition of the scalar form factors}  
\label{sec:sff}  
One can define an extended QCD Lagrangian allowing for the presence of external sources.  
In this way the identification of matrix elements of quark currents can be  
done easily. For instance, a scalar source can be added simply as:  
\be  
\label{ssour}  
-\bar{q}\, \Sigma\, {q}~,  
\ee   
where $q$ embodies the three light quarks,  $u$, $d$ and $s$. The QCD  
current quark mass term can be obtained from such a scalar source by  
setting,  
\be  
\label{mqcd}  
\Sigma= {\rm diag}(m_u,m_d,m_s)~.  
\ee  
This is the standard method of treating explicit chiral symmetry  
breaking in CHPT (or any similar effective field  
theory). Consequently, we can work out the scalar quark--antiquark  
operators,  
\ba  
\label{ss}  
\bar{u}u&=& -\frac{\partial {\mathcal{L}}_{\rm QCD}}{\partial \Sigma_{11}}   
          = -\frac{\partial {\mathcal{L}}_{\rm QCD}}{\partial m_u}~, \nonumber \\  
\bar{d}d&=&- \frac{\partial {\mathcal{L}}_{\rm QCD}}{\partial \Sigma_{22}}  
          = -\frac{\partial {\mathcal{L}}_{\rm QCD}}{\partial m_d}~, \nonumber \\  
\bar{s}s&=&- \frac{\partial {\mathcal{L}}_{\rm QCD}}{\partial \Sigma_{33}}  
          = -\frac{\partial {\mathcal{L}}_{\rm QCD}}{\partial m_s}~.  
\ea  
One can include in the same way as in the QCD Lagrangian the external sources in   
the effective CHPT Lagrangian~\cite{gl} simply based on symmetry arguments.   
In the lowest order chiral effective Lagrangian, ${\mathcal{L}}_2$, the scalar  
source appears in the mass term  
\be  
\label{sscpt}  
{\cal L}_{2}^{\rm mass} = \frac{1}{4} f^2 \langle U^\dagger \chi+\chi^\dagger U \rangle~,  
\ee  
with $f$ the meson decay constant (in the chiral limit), $\chi\equiv 2  
B_0 \Sigma$ and  $B_0$ is a constant not fixed by symmetry. This  
constant parameterizes the strength of the quark--antiquark condensation  
in the non--perturbative vacuum, $B_0 = |\langle 0 | \bar{q}q|0\rangle | 
/ f^2$. The trace in flavor space is denoted by $\langle \ldots \rangle$.  
The octet of Goldstone bosons is collected in the  
matrix--valued unimodular field $U(x)$,   
\be  
\label{u}  
U=\exp\left( \frac{i\sqrt{2}}{f}\Phi \right)  
\ee  
with   
\be  
\label{Phi}  
\Phi = \left[ \matrix{\frac{1}{\sqrt{2}}\pi^0+\frac{1}{\sqrt{6}}\eta_8 &   
\pi^+ & K^+ \cr \pi^- & -\frac{1}{\sqrt{2}}\pi^0+\frac{1}{\sqrt{6}}\eta_8 & K^0   
\cr K^- & {\bar K^0} & -\frac{2}{\sqrt{6}}\eta_8} \right]~.  
\ee  
It is then straightforward to work out the scalar--isoscalar  
quark-antiquark operators from the effective Lagrangian,  
\ba  
\label{ssl2}  
\bar{u}u&=&-\frac{\partial{\mathcal{L}}_2 }{\partial \Sigma_{11}}=-f^2 B_0  
\left[1-\frac{1}{f^2}\left(\pi^+ \pi^- + K^+ K^- + \frac{(\pi^0)^2}{2} + \frac{\eta_8^2}{6} +   
\frac{\pi^0 \eta_8}{\sqrt{3}}\right)+...\right] \nonumber \\  
\bar{d}d&=&-\frac{\partial{\mathcal{L}}_2 }{\partial \Sigma_{22}}=-f^2 B_0  
\left[1-\frac{1}{f^2}\left(\pi^+ \pi^- + K^0 \overline{K}^0 + \frac{(\pi^0)^2}{2} +   
\frac{\eta_8^2}{6} - \frac{\pi^0 \eta_8}{\sqrt{3}}\right)+...\right] \nonumber \\  
\bar{s}s&=&-\frac{\partial {\mathcal{L}}_2}{\partial \Sigma_{33}}=-f^2  
B_0\left[1-\frac{1}{f^2}\left(K^+K^- + K^0 \overline{K}^0 +\frac{2}{3}\eta_8^2\right)+...  
 \right]  
\ea  
where the ellipsis denotes terms of higher order in the meson fields  
not needed here. From the last of these equations one concludes that the  
strangeness component of the pion should be small since it only comes
in at higher orders. From this representation of the scalar  
operators, one can deduce the pertinent expressions for the scalar form  
factors. For our purposes, it is sufficient to consider pure isospin zero  
($I=0$) states formed from a pion or kaon--anti-kaon pair, i.e.  
\ba  
\label{i0state}  
|\pi\pi\rangle&=&\frac{1}{\sqrt{6}}|\pi^+\pi^- + \pi^-\pi^+ + \pi^0  
 \pi^0\rangle~,  
 \nonumber \\  
|K\bar{K}\rangle &=& \frac{1}{\sqrt{2}}|K^+ K^- + K^0 \overline{K}^0\rangle~.  
\ea  
Note the extra factor $1/\sqrt{2}$ in the definition of the $I=0$  
 $|\pi\pi\rangle$ state, it is introduced to take care that in the isospin   
basis states the pions behaves as identical particles.   
Combining this with Eq.(\ref{ssl2}), one can easily calculate the lowest order  
(tree level) CHPT results (remember the normalization of the non--strange  
quark operator given in Eq.(\ref{nquarks})),  
\ba  
\label{rest}  
\langle0|\bar{n}n|\pi\pi\rangle&=&\sqrt{3}\,B_0~, \nonumber \\  
\langle0|\bar{n}n|K\overline{K}\rangle&=& B_0~,  \nonumber \\  
\langle0|\bar{s}s|\pi\pi\rangle&=&0 ~,\nonumber \\  
\langle0|\bar{s}s|K\overline{K}\rangle&=&\sqrt{2}\,B_0~.   
\ea  
As anticipated, to leading order the two--pion system has no strangeness  
component. To all orders, these matrix elements are given in terms of  
four {\sl scalar} form factors,\footnote{We remark that more commonly the
definition of these form factors includes the pertinent quark masses, such
that e.g. the non-strange scalar form factor of the pion is defined
via $ \langle0|\hat{m} \, (\bar{u}u + \bar{d}d)|\pi\pi\rangle = M_\pi^2\,
\Gamma_\pi(s)$. For our later discussion, the overall normalization does not
play a role but should be kept in mind.}  
\ba  
\label{rest2}  
\langle0|\bar{n}n|\pi\pi\rangle&=&\sqrt{2}\,B_0\, \Gamma^n_1(s)~, \nonumber \\  
\langle0|\bar{n}n|K\overline{K}\rangle&=&\sqrt{2}\, B_0 \, \Gamma^n_2(s)~, \nonumber \\  
\langle0|\bar{s}s|\pi\pi\rangle&=&\sqrt{2} B_0 \, \Gamma^s_1(s) ~,\nonumber \\  
\langle0|\bar{s}s|K\overline{K}\rangle&=&\sqrt{2}\,B_0 \, \Gamma^s_2(s)~.  
\ea  
Here, the following notation is employed. The superscript $s/n$ refers to the  
strange/non--strange quark operator whereas the subscript $1,2$ denotes pions  
and kaons, respectively. In the following we will remove from  
Eqs.(\ref{rest},\ref{rest2})  the overall factor $\sqrt{2}   
B_0$, since the experimental data on the $J/\Psi \to \phi M\bar M$ decays are  
not normalized.  
  
\subsection{Next-to-leading order pion and kaon scalar form factors}

\begin{figure}[htb]
\vskip 0.5cm
\centerline{
\epsfysize=1.2 in 
\epsffile{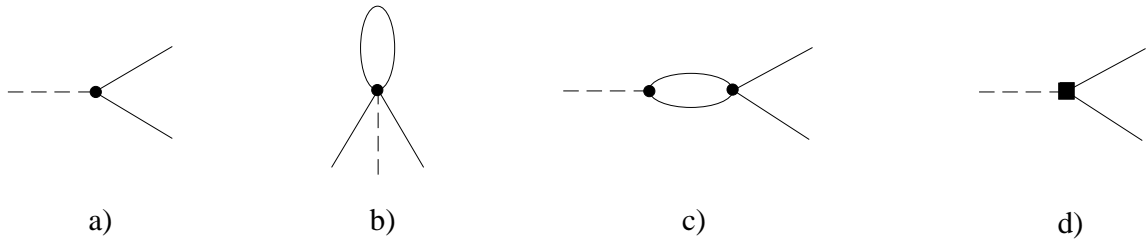}
}
\vskip 1cm 

\caption[]{Feynman diagrams for the calculation of the scalar form factors
at leading and next-to-leading order in CHPT.
The scalar source is indicated by the dashed line and the solid
lines refer to the psuedoscalars (pions and kaons). From left to right: 
a) Lowest order, b) tadpole contributions, c) unitarity contributions and d)
local contact terms
from the ${\mathcal{O}}(p^4)$ CHPT Lagrangian. The full circles indicates that the
vertices come from the lowest order CHPT Lagrangian and the full
squares symbolize an insertion from  next-to-leading order. The wave function
renormalization diagrams are not drawn.}\label{fig:dia}
\end{figure}

The pion scalar form factors $\Gamma^n_1(s)$ and $\Gamma^s_1(s)$ were 
calculated in Ref.\cite{gl} up to one loop in CHPT. 
Since they were not explicitly given in Ref.\cite{gl}, we give here
the pertinent expressions:
\ba
\label{glpi}
\Gamma^n_1(s)&=&\sqrt{\frac{3}{2}}\left\{1+\mu_\pi-\frac{1}{3}\mu_\eta+\frac{16 m_\pi^2}{f^2}
\left(2L_8^r-L_5^r\right)+8(2 L_6^r -L_4^r)\frac{2
m_K^2+3m_\pi^2}{f^2}+f(s)+\frac{2}{3}\,\widetilde{f}(s)\right\}~, \nonumber \\
\Gamma^s_1(s)&=&(2 L_6^r -L_4^r )\frac{8\sqrt{3}\, m_\pi^2}{f^2}+
\frac{1}{\sqrt{3}}\widetilde{f}(s)~, 
\ea 
with $f(s)$ and  $\widetilde{f}(s)$  given by
\ba
\label{ft}
f(s)&=&\frac{2s-m_\pi^2}{2f^2}\bar{J}_{\pi\pi}(s)-\frac{s}{4f^2}\bar{J}_{KK}(s)-
\frac{m_\pi^2}{6f^2}\bar{J}_{\eta\eta}(s)+
\frac{4s}{f^2}\left\{L_5^r-\frac{1}{256\pi^2}\left(4\log
\frac{m_\pi^2}{\mu^2}-\log\frac{m_K^2}{\mu^2}+3 \right) \right\}~,\nonumber \\
\widetilde{f}(s)&=&\frac{3}{4}\,\frac{s}{f^2}\bar{J}_{KK}(s)+\frac{m_\pi^2}{3f^2}
\bar{J}_{\eta\eta}(s)+\frac{12s}{f^2}\left\{L_4^r-\frac{1}{256 \pi^2}\left(\log
\frac{m_K^2}{\mu^2}+1 \right)\right\}~, 
\ea
and $\bar{J}_{PP}(s)$ ($P=\pi,K,\eta$) is the standard meson loop function~\cite{gl}
\be
\label{jb}
\bar{J}_{PP}(s)=\frac{1}{16 \pi^2}\left(2+ \sigma_P (s)
\log \frac{\sigma_P(s)-1}{\sigma_P(s)+1} \right)~,
\ee
and $\mu$ is the scale of dimensional regularization.
The quantities $\mu_P$ in Eq.(\ref{glpi}) are given by
\be
\mu_P=\frac{m_P^2}{32\pi^2 f^2}\log\frac{m_P^2}{\mu^2}~.
\ee
The scalar kaon form factors at next-to-leading order
in CHPT are not given explicitly in the literature. 
We fill here this gap by performing such a calculation. This implies calculating 
the diagrams shown in Fig.\ref{fig:dia}, which comprise the lowest order CHPT
result, Fig.\ref{fig:dia}a, already derived in section~\ref{sec:sff}, 
the tadpole contribution, Fig.\ref{fig:dia}b, and the
unitarity corrections, Fig.\ref{fig:dia}c. 
The vertices for these diagrams come from the lowest
order CHPT Lagrangian. We note that wave function renormalization
diagrams are not depicted in this figure. 
Finally, in Fig.\ref{fig:dia}d, the local contribution coming from the
${\mathcal{O}}(p^4)$ CHPT Lagrangian is depicted. These terms are parameterized
in terms of the scale--dependent, renormalized low energy constants
$L_i^r (\mu)$ (in our case $i=4,5,6,8$). Evaluating these diagrams leads to
\ba
\label{kksf}
\Gamma^n_2(s)&=& \frac{1}{\sqrt{2}} \Bigg\{ 1+\frac{4 L_5^r}{f^2}(s-4m_K^2)+
\frac{8L^r_4}{f^2}(2s-6m_K^2-m_\pi^2)+L_8^r\frac{32 m_K^2}{f^2}+
\frac{16 L^r_6}{f^2}(6 m_K^2+m_\pi^2)+\frac{2}{3}\mu_\eta\nonumber \\
&+&\frac{9s-8m_K^2}{36 f^2}J^r_{\eta\eta}(s)+
\frac{3s}{4f^2}\left[J^r_{\pi\pi}(s)+J^r_{KK}(s)\right] \Bigg\}~, \nonumber \\
\Gamma^s_2(s)&=&1+\frac{4 L_5^r}{f^2}(s-4m_K^2)+\frac{8L^r_4}{f^2}(s-4
m_K^2-m_\pi^2)+L_8^r\,\frac{32 m_K^2}{f^2} +\frac{16 L_6^r}{f^2}(4m_K^2+m_\pi^2)+
\frac{2}{3}\mu_\eta\nonumber \\ &+&\frac{9s-8m_K^2}{18 f^2}J^r_{\eta\eta}(s)+
\frac{3s}{4f^2}J^r_{KK}(s)~.
\ea 
As a test of our calculations we have checked that the infinities,
associated with the wave function renormalization contributions and the loops in 
Fig.\ref{fig:dia}c and \ref{fig:dia}d are properly absorbed by the
infinite parts of the pertinent low energy constants and thus the expressions given
in Eqs.(\ref{kksf}) are finite.

\subsection{Unitarity requirements}  
  
We now discuss the constraints that unitarity imposes on the scalar form  
factors. Of course, at low energies, one can simply work with CHPT and  
treat unitarity in a perturbative fashion. Here, however, we are interested  
also at energies of the order of 1~GeV, which requires some resummation  
technique and also the channel coupling between the $\pi\pi$ and the  
$K\overline{K}$ systems has to be taken into account. This has been   
elaborated in big detail in Ref.\cite{npa} and we present here the  
formalism necessary to discuss the scalar form factors. For convenience,  
we employ the matrix notation already introduced in the previous  
subsection, i.e.  pions are labeled by the index 1 and kaons by the index 2.  
>From Ref.\cite{npa}, we have the following expression for the $T$-matrix  
for meson--meson scattering,  
\be  
\label{t}  
T(s)=\left[I+K(s)\cdot g(s) \right]^{-1}\cdot K(s)  ~,
\ee  
where $s$ denotes the centre-of-mass energy squared and $K(s)$ can be obtained    
from the lowest order CHPT Lagrangian,   
\ba  
\label{k0}  
K(s)_{11}&=&\frac{s-m_\pi^2/2}{f_\pi^2}~,\nonumber \\  
K(s)_{12}&=&\frac{\sqrt{3}s}{4 f_\pi^2}~,\nonumber\\  
K(s)_{22}&=&\frac{3 s}{4 f_\pi^2}~.  
\ea  
We remark that because of time reversal, both $K(s)$ and $T(s)$ are symmetric  
functions, so that $K(s)_{21} = K(s)_{12}$ and similarly for $T(s)$.  
The matrix $g(s)$ is also diagonal and given by~\cite{prd}  
\be  
\label{gs}  
g(s)_{i}=\frac{1}{16 \pi^2}\left\{ \sigma_i(s) \log \frac  
{\sigma_i(s) \sqrt{1+\frac{m_i^2}{q_{\rm max}^2}}+1}  
{\sigma_i(s) \sqrt{1+\frac{m_i^2}{q_{\rm max}^2}}-1}-2 \log \left[  
\frac{q_{\rm max}}{m_i}\left(1+\sqrt{1+\frac{m_i^2}{q_{\rm max}^2}} \right)  
\right] \right\}~, \,\,\,\, i = 1,2~,  
\ee  
where $\sigma_i(s)=\sqrt{1-4m_i^2/s}$, $f_\pi\simeq 93$ MeV is the weak pion decay   
constant, $m_i$ are the masses of the pions ($m_1=138$ MeV) and kaons ($m_2=495.7$ MeV) and  
$q_{\rm max}=0.9$ GeV is a cut-off in three-momentum space. 
On the other hand, $g(s)_i$ can also be calculated in dimensional
regularization, using the standard $\overline{MS}-1$ scheme employed in
CHPT~\cite{gl},
\be\label{gis}
g(s)_i = {1\over (4\pi)^2} \biggl( -1 + \log{m_i^2\over \mu^2} +
\sigma_i (s) \log{ \sigma_i (s) + 1\over \sigma_i (s) -1} \biggr)
= - J^r_{ii} (s)~,
\ee
where the last equality follows from the definition of the renormalized two--meson
loop function~\cite{gl}
\be\label{gis17}
\bar{J}_{ii}(s) - \frac{1}{16 \pi^2}\left(1+\log\frac{m_i^2}{\mu^2} \right)=
J^r_{ii}(s)~.
\ee
Note that we have changed the subscript ``$PP\,$'' appearing in
Eq.(\ref{jb}) into ``$ii\,$'' to conform with our matrix notation. 
By expanding
Eq.(\ref{gs}) in terms of $m_i/q_{\rm max}$ one can easily see, as
discussed in appendix~2 of Ref.\cite{prd}, that the differences between 
$g(s)_i$ in Eq.(\ref{gs}) and Eq.(\ref{gis})
are of higher order in the chiral expansion, i.e. of order ${\cal
  O}(m_i^2/q^2_{\rm max})$, for the following value of the scale
$\mu$,
\be\label{mumu}
\mu = {2 q_{\rm max} \over \sqrt{e}} \simeq 1.2 \, q_{\rm max}~.
\ee
For energies above the threshold of the state $i$, unitarity implies the following  
relation between form factors and the $I=0$ $T$-matrix:  
\be  
\label{uni}  
\hbox{Im}~\Gamma_i(s)=\sum_j  \Gamma_j(s)  
\frac{p_j(s)}{8\pi\sqrt{s}}\theta(s-4m_j^2)(T_{ji}^{\hbox{S-wave}}(s))^*  
\ee  
with $p_i(s)=\sqrt{s/4-m_i^2}$ the modulus of the c.m.  
three-momentum of the state $i$, and the strong amplitudes are projected on the S-wave.  
In the former equation we have suppressed the superscript ``$n$'' or ``$s$'', appearing in 
Eq.(\ref{rest2}), since the previous equation applies to any of them. Finally, in what follows,
we will also remove the superscript ``S-wave'' with the understanding that any partial wave is  
projected onto the S-wave. Taking now the complex conjugate on the right-hand-side of 
Eq.(\ref{uni}) and using the fact   
that the $T$-matrix is symmetric, we can rewrite Eq.(\ref{uni}) in matrix notation as:  
\be  
\label{uni2}  
\hbox{Im}~\Gamma(s)=T(s)\cdot \frac{Q(s)}{8\pi\sqrt{s}}\cdot \Gamma^*(s)  
\ee  
where
\begin{equation}
Q(s)=\left(\matrix{p_1(s)\theta(s-4m_1^2) & 0 \cr 0 & p_2(s)\theta(s-4m_2^2)}\right)~,
\quad \Gamma(s)=\left(\matrix{\Gamma_1(s) \cr \Gamma_2(s)}\right)~.
\end{equation}
Substituting in the previous equation $\hbox{Im}~\Gamma(s)$ by  
$(\Gamma(s)-\Gamma(s)^*)/(2i)$ and $T(s)$ by its expression given in Eq.(\ref{t}), one has:  
\ba  
\label{uni3}  
\Gamma(s)&=&\left\{ I+\left[ I+K(s)\cdot g(s) \right]^{-1}\cdot K(s)\cdot i  
\frac{Q(s)}{4 \pi\sqrt{s}} \right\}\cdot \Gamma(s)^* \nonumber \\  
&=&\left[I+K(s)\cdot g(s) \right]^{-1}\cdot \left\{ I+K(s)\cdot g(s) + K(s)   
\cdot i\frac{Q(s)}{4\pi\sqrt{s}}\right\}\cdot \Gamma(s)^*  ~.
\ea  
Taking into account that the $K(s)$-matrix, Eq. (\ref{k0}), is real and that   
\be  
\label{g*}  
g(s)^*=g(s)+i\frac{Q(s)}{4\pi\sqrt{s}}  
\ee  
since  
\be  
\label{g*2}  
\hbox{Im}~g(s)=-\frac{Q(s)}{8\pi\sqrt{s}}  
\ee  
we can write Eq.(\ref{uni3}) as:  
\be  
\label{uni4}  
\left[I+K(s)\cdot g(s) \right]\cdot \Gamma(s)=\left[I+K(s)\cdot g(s)^* \right]\cdot  
\Gamma(s)^*  ~.
\ee  
This tells us that the quantity $\left[I+K(s)\cdot g(s) \right]\cdot \Gamma(s)$ has no cuts  
since the only one which appears in $g(s)$ and $\Gamma(s)$, the right or unitarity cut, is  
removed. Therefore, we can express $\Gamma(s)$ as:  
\be  
\label{gamma}  
\Gamma(s)=\left[I+K(s)\cdot g(s)\right]^{-1} \cdot R(s)  
\ee   
with $R(s)$ being a vector of functions free of any singularity. We remark that
this procedure of taking into account the final state interactions is based on 
the work presented in Ref.\cite{basde}. In the following, we will fix  $R(s)$ by requiring the matching of 
Eq.(\ref{gamma}) to the
next-to-leading order (one loop) CHPT $\pi\pi$ and $K\bar{K}$ scalar form factors. 
These are calculated in the next subsection. 

It is worth to stress that Eq.(\ref{gamma}), given in terms of a
vector of functions $R(s)$ without any cut, can be applied to any K-matrix 
without unphysical cut contributions, as the one derived in ref. \cite{npa}. The use of 
the strong amplitudes calculated from this reference is appealing for several reasons: 
1) Because of their simplicity, 2) they have  been already successfully used to describe
 many two meson production processes, as discussed in the introduction, and 3) higher 
order corrections to the kernel used in ref. \cite{npa} are not necessary to match with 
the next-to-leading order CHPT scalar form factors. In fact, 
in ref. \cite{nd} one can find a detailed comparison between the
approach of 
ref.\cite{npa} and the more general ones described in refs.\cite{nd} and \cite{prd}. The
 main conclusion is that, apart from the detail of including  (or not) the 
$\eta\eta$ channel as already discussed, the unitarity corrections coming from the 
rescattering of the lowest order CHPT kernel completely dominate the strong S-wave 
$I=0$ scattering amplitudes up to about 1.2 GeV. Thus, one would not expect relevant 
departures from the use of the strong amplitudes from ref.\cite{npa}
or from refs.\cite{prd} or \cite{nd}. In fact, all these approaches
give rise to very similar pole
 positions for the $f_0(980)$ and $\sigma$ mesons.  For higher energies new effects have 
to be taken into account as e.g. the contributions from a pre-existing octet of scalar 
resonances around 1.4 GeV~\cite{nd}\footnote{Preexisting means here that these 
resonances with a mass around 1.4 GeV are as ``elementary'' as the basic fields $\pi$, 
$K$ or $\eta$.} and the increasingly important role played by multiparticle states, 
basically the $4\pi$ intermediate state. In addition, one has to deal with more relevant
interaction vertices between the various fields than those given in Eq.(\ref{lagS})
as discussed in section~\ref{sec:model}.

\subsection{Matching with chiral perturbation theory}
\label{sec:match}
The general expression for the scalar form factors given in Eq.(\ref{gamma}) can be further
constrained by matching it to the one loop CHPT expression given in Eqs.(\ref{glpi}, 
\ref{kksf}).
This ensures that for energies where CHPT is applicable, these form factors fulfill
all requirements  given by chiral symmetry and the underlying power counting. This
matching procedure essentially fixes the vector $R(s)$. We remark that since in our 
unitarization procedure we are not considering the $\eta\eta$ channel we thus can not 
reproduce the chiral logarithms associated with this channel. Therefore, we will only 
consider the contribution form this channel to the value of the form factors at $s=0$ 
and we will not include any $s$ dependence. This approximation should not induce any
sizeable theoretical error because the influence of the $\eta\eta$ channel was found
to be significant only above its threshold \cite{nd,paquito} as already discussed in the
introduction (when comparing the results of Ref.\cite{npa} and Ref.\cite{nd}).

\medskip\noindent  
We only discuss in detail the matching for the form factor $\Gamma^s_2(s)$. 
The procedure for the  other form factors is completely analogous and we thus only give 
the
final results for them. From Eq.(\ref{gamma}) one has:
\be
\label{m1}
\Gamma^s (s)=\left[I+K(s)\cdot g(s) \right]^{-1}\cdot R^s(s)=\left[I-K(s)\cdot g(s)
\right]\cdot R^s(s)+{\mathcal{O}}(p^4)
\ee
where the superscript ``$s$'' in $R^s(s)$ indicates that we are considering the $\bar{s}s$ form
factor. From the former equation and Eq.(\ref{rest}) one has that 
$R^s(s)_1={\mathcal{O}}(p^2)$ and that $R^s(s)_2=1+{\mathcal{O}}(p^2)$. Hence, we can
recast Eq. (\ref{m1}) as:

\be
\label{m2}
\Gamma^s(s)_2=R^s(s)_2-K(s)_{22}\,g(s)_{2}+{\mathcal{O}}(p^4)=
R^s(s)_2+\frac{3s}{4f^2}\,J^r_{KK} (s)+{\mathcal{O}}(p^4)
\ee  
at the regularization scale $\mu=1.2 q_{\rm max}$.
Comparing this result with the one given in Eq.(\ref{kksf}) leads to
\ba
\label{rs2}
R^s(s)_2&=&1+\frac{4\,L^r_5}{f^2}(s-4m_K^2)+\frac{8L^r_4}{f^2}(s-4m_K^2-m_\pi^2)+L^r_8
\frac{32 m_K^2}{f^2}+\frac{16
L^r_6}{f^2}(4m_K^2+m_\pi^2)+\frac{2}{3}\mu_\eta \nonumber \\ 
&+&\frac{m_K^2}{36\,\pi^2\,f^2}(1+\log \frac{m_\eta^2}{\mu^2})~,
\ea
using the Gell-Mann--Okubo relation $m^2_\eta=4 m_K^2/3-m_\pi^2/3$, the deviations from it
being of higher order for our purpose. 
Proceeding in an analogous way for the other form factors one concludes:
\ba
\label{R}
R^n(s)_1&=&\sqrt{\frac{3}{2}}\Bigg\{1+\frac{4(L_5^r+2
L_4^r)}{f^2}s+\frac{16
(2L_8^r-L_5)}{f^2}m_\pi^2+\frac{8(2L_6^r-L_4^r)}{f^2}(2m_K^2+3m_\pi^2)\nonumber \\ 
&-&\frac{m_\pi^2}{32\,\pi^2\,f^2}-\frac{1}{3}\mu_\eta\Bigg\} ~, \nonumber \\
R^n(s)_2&=&\frac{1}{\sqrt{2}}\Bigg\{1+\frac{4 L_5^r}{f^2}(s-4m_K^2)+\frac{8L^r_4}{f^2}(2s-6
m_K^2-m_\pi^2)+L_8^r\,\frac{32 m_K^2}{f^2} +\frac{16 L_6^r}{f^2}(6m_K^2+m_\pi^2)+
\frac{2}{3}\mu_\eta \nonumber \\
&+&\frac{m_K^2}{72\,\pi^2\,f^2}(1+\log\frac{m_\eta^2}{\mu^2})\Bigg\} ~, \nonumber \\
R^s(s)_1&=&\sqrt{3}\, \Bigg\{ \frac{4
  L_4^r}{f^2}(s-2m_\pi^2)+\frac{16 L_6^r}{f^2}m_\pi^2\Bigg\}~.
\ea
Notice that $R^s(s)_1$ is subleading in large $N_c$, i.e. of
${\mathcal{O}}(N_c^{-1})$, while the other quantities  in Eqs.(\ref{rs2},\ref{R}) are
of order ${\mathcal{O}}(1)$ in this counting. This is expected since the production of pions from
an $\bar{s}s$ source is  subleading in 
large $N_c$ QCD~\cite{witten}. We also see, as already stressed in section~\ref{sec:sff}
that $R^s(s)_1$ is ${\mathcal{O}}(p^2)$ in the chiral counting.
Once the functions $R^{n,s}(s)$  have been determined, the final
expressions for the form factors are obtained by making use of Eq.(\ref{gamma}, \ref{k0},
\ref{gs}). Finally, one has to take into account that, when using Eqs.(\ref{rs2},
\ref{R}), the regularization scale is $\mu=1.2 q_{\rm max}\simeq 1.08$ GeV. Therefore, 
we have to run the low energy constants $L_i^r(\mu)$ to this 
scale from the usual ones $\mu=m_\eta$ or $\mu=m_\rho$, with $m_\eta$, $m_\rho$ the mass 
of the $\eta$, $\rho$ meson, respectively, by using the appropriate
$\beta$--functions given in Ref.\cite{gl}.

\section{Results}  
\label{sec:res}
We will first discuss the results for the $J/\Psi \rightarrow \phi \pi\pi(K\bar{K})$ decays 
and then we will also consider to some extent the $J/\Psi \rightarrow \omega \pi\pi$ decay.
To be more specific, we consider the S--wave contribution to these decay modes.

\subsection{The $\phi$-meson case}

Considering the phase space of three particles \cite{pdg} we can write the unpolarized   
event distribution for the $J/\Psi \rightarrow \phi \pi^+\pi^- (K^+K^-)$ reactions as:  
\ba  
\label{dis}  
\frac{d\,N(W)_i}{d\,W}&=&\frac{{\mathcal{C}}_\phi^2\,G_i^2}{(2\pi)^3 12\, 
m_{J/\Psi}^2}\,|\Gamma^s_i(s)+\lambda_\phi \,  
\Gamma^n_i(s)|^2\left[1+\frac{(m_{J/\Psi}^2+m_\phi^2-W^2)^2}{8 m_\phi^2 m_{J/\Psi}^2}  
\right]\nonumber \\  
&\times&\sqrt{[W^2-4m_i^2]\left[(m_{J/\Psi}^2-W^2-m_\phi)^2-4 m_\phi^2 W^2\right]},  
\ea  
where $i=1$ refers to the $\pi^+\pi^-$ and $i=2$ to the $K^+ K^-$
system, in order. Furthermore, $W$ is the total 
energy in the c.m. of the two pions or kaons, $G_i$ is basically a Clebsch-Gordan 
coefficient equal to 4/3 for pions and 1/2 for kaons, respectively, and ${\mathcal{C}}_\phi$ a  
normalization constant depending on the experiment, in our case DM2~\cite{dm2} 
or MARK-III~\cite{mk3}. In comparing with the experimental data, we will
average Eq.(\ref{dis}) over the width of the  bin (as given by the
corresponding experiment).
As discussed in section~\ref{sec:match}, 
our calculated form factors depend on the CHPT low energy
constants $L_4^r$, $L_5^r$, $L_6^r$ and $L_8^r$. From these, only $L_5^r$ and $L_8^r$ are
relatively well determined. Their most recent  values, given in Ref.\cite{bij} from an
 ${\mathcal{O}}(p^6)$ CHPT analysis of the $K_{\ell\,4}$ form factors, are:
\be
\label{bij0}
10^3\, L_5^r(M_\rho)= 0.65 \pm 0.12~, \;\;\; 10^3 L_8^r(M_\rho)=0.48 \pm 0.18.
\ee 
At the scale $\mu=1.2 q_{\rm max}\simeq 1.08$ GeV they are:
\be
\label{bij}
10^3 \, L_5^r (1.08~{\rm GeV}) = -0.15\pm 0.12 ~, \;\;\; 
10^3 \, L_8^r (1.08~{\rm GeV}) = 0.26\pm 0.18~,
\ee
On the other hand, $L_4^r$ and $L_6^r$ are only poorly known and their present 
values~\cite{gl} can be considered as stemming more from an estimation of
their order of magnitude than from a truly 
phenomenological fit. According to Ref.\cite{gl}, it is estimated that for a
regularization scale $\mu$ between 0.5 and 1.0 GeV one should have
$10^3\,L_4^r\simeq \pm0.5$ and $10^3\,L_6\simeq \pm 0.3$. A more
recent determination~\cite{daphne} gives $10^3\,L_4^r (m_\rho) =
-0.3\pm 0.5$ and $10^3\,L_6^r (m_\rho) = -0.2 \pm 0.3$ so that at
a scale of 1.08~GeV one has 
\be\label{est}
10^3 \, L_4^r (1.08~{\rm GeV}) = -0.57 \pm 0.5~, \quad 
10^3 \, L_6^r (1.08~{\rm GeV}) = -0.36 \pm 0.3~. 
\ee
Again, this estimate relies on OZI (large $N_c$) arguments. To be more
precise, one sets $L_{4,6}^r$ to zero at the scale $\mu = m_\eta$, which
is of course somewhat arbitrary. One can also make use of information about the 
low energy coupling constant $\ell_4^r$ coming from two flavor CHPT by
means of the  relation \cite{gl}, $\ell_4^r (\mu ) = 
8L_4^r (\mu) + 4L_5^r (\mu) - \nu_K/2  + {\cal O}(p^6)$ with $\nu_K =
[\ln(M_K^2/\mu^2)+1]/ 32 \pi^2$. The low energy coupling constant $\ell_4^r$ has been 
determined at ${\mathcal{O}}(p^4)$ in ref. \cite{gl} with the result $10^3\,\ell_4^r=1.2
 \pm 6$ at the scale $\mu=1.08$ GeV. However, in ref. \cite{bij98} making use of the 
analytically deduced pion 
scalar form factor $\bar{u}u+\bar{d}d$ up to and including ${\mathcal{O}}(p^6)$, they 
update the previous value and give the improved result $10^3\,\ell_4^r=1.8\pm 1.9$ at 
the same scale. Although the central value from both determinations is very similar the 
error is much smaller in the second case. With the value for $L_5^r$ given in Eq.
(\ref{bij}), we obtain $10^3 \,L_4^r (1.08~{\rm GeV}) = 0.1 \pm 0.7$ when using 
$\ell_4^r$ from ref. \cite{gl} and 
\begin{equation}
\label{bij98}
10^3\,L_4^r(1.08~{\rm GeV})=0.19 \pm 0.25
\end{equation}
 in the case of ${\mathcal{O}}(p^6)$ SU(2) CHPT \cite{bij98}. The 
value obtained in this way for $L_4^r$ is also consistent with zero, but on the 
positive side, in stark contrast to the value given in Eq.(\ref{est}) whose central 
value is very far from the more precise determination coming from refs. 
\cite{bij,bij98} as given in Eq.(\ref{bij98}). 
Thus, although at present no
precise determination of this low energy constant is available, rather strong 
constraints on its value can be obtained by combining the determination of the low 
energy constants making use of ${\mathcal{O}}(p^6)$ CHPT, both in its SU(2) 
\cite{bij98} and SU(3) \cite{bij} forms. 

\medskip\noindent
More recently, some constraints on the couplings $L_4^r$ and $L_6^r$
have been reported~\cite{mouss,mouss2,stern}. The determination of $L_4^r$ relies on 
a comparison of the CHPT series at next-\cite{mouss} and 
next-to-next-leading
 order \cite{mouss2} with a phenomenologically determined scalar form factor. In ref. 
\cite{mouss} the value $10^3\,L_4^r(1.08~\rm{GeV})\simeq 0.14$ is given without errors 
and the band of values $-0.12\leq 10^3\, L_4^r(1.08~\rm{GeV})<0.04$ is reported in 
ref. \cite{mouss2}. We note that the second set of values for $L_4^r$ \cite{mouss2} is 
in the lower limit of
the value for $L^r_4$ given in Eq.(\ref{bij98}).
 Nevertheless, the determination of $L_6^r$ \cite{mouss,mouss2} is not so well sounded 
due to strong simplifying working assumptions when 
computing phenomenologically the quark correlator $(\bar{u}u+\bar{d}d)\bar{s}s$ 
\cite{mouss,mouss2}. It is also stated in ref.\cite{mouss} that the positivity of the 
fermionic measure gives rise to a lower bound for $L_6^r$, $10^3L_6^r (1.08~\rm{GeV})
\ge 0.03$. However, this bound is somewhat arbitrary since the only necessary requirement 
to make use of the positivity of the fermionic measure is that all the three light 
quark masses have to be equal. In ref.\cite{mouss}, they were set equal to the strange 
quark mass, but they could as well have been taken on another
value. In fact, for the average light quark masses $(m_u+m_d)/2$,
the corresponding lower bound is: $10^3
L_6^r(1.08~\rm{GeV})\geq-0.75$. The bound based on using the strange
quark mass can only be maintained if one assumes the next--to--leading
order corrections to be of canonical size, $c M_K^2/(4\pi f_\pi)^2$, with
$c$ a number of order one.\footnote{We are grateful to Bachir
Moussallam for a claryfing discussion on this topic.}

\begin{figure}[htb]
\centerline{
\epsfysize=3.9in 
\epsffile{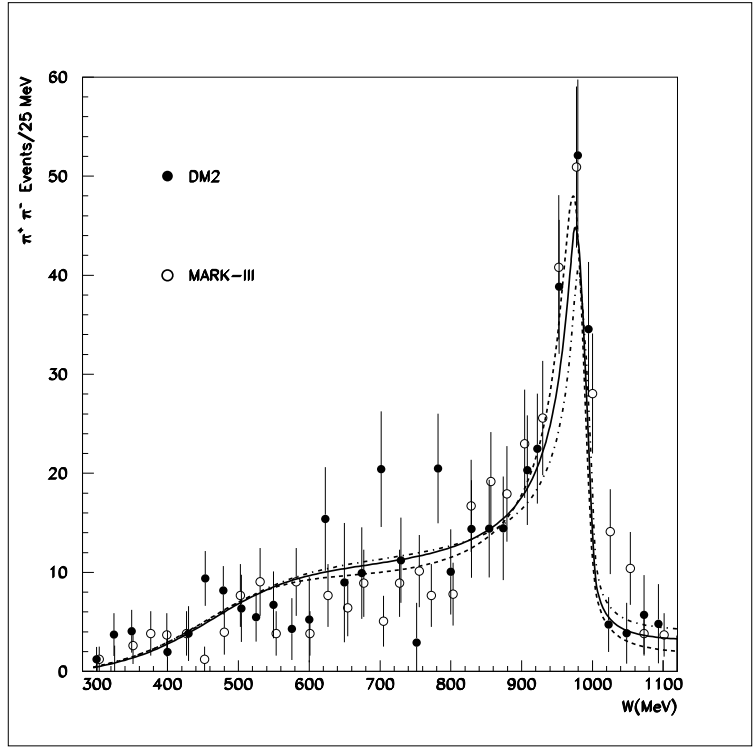}}
\vskip 0.4cm 
\caption[]{$\pi^+\pi^-$ event distribution in the $J/\Psi \rightarrow \phi \pi^+ \pi^-$
decay. The width of the bin is 25 MeV. The solid line corresponds to the fit 
Eq.(\ref{result}) with $q_{\rm max}=0.9\,$GeV. The dashed line is the best fit with 
$q_{\rm max}=1\,$GeV and analogously the dashed--dotted line for $q_{\rm max}=0.8\,$GeV.}
\label{fig:25}
\end{figure}

\begin{figure}[htb]
\centerline{
\epsfysize=3.9in 
\epsffile{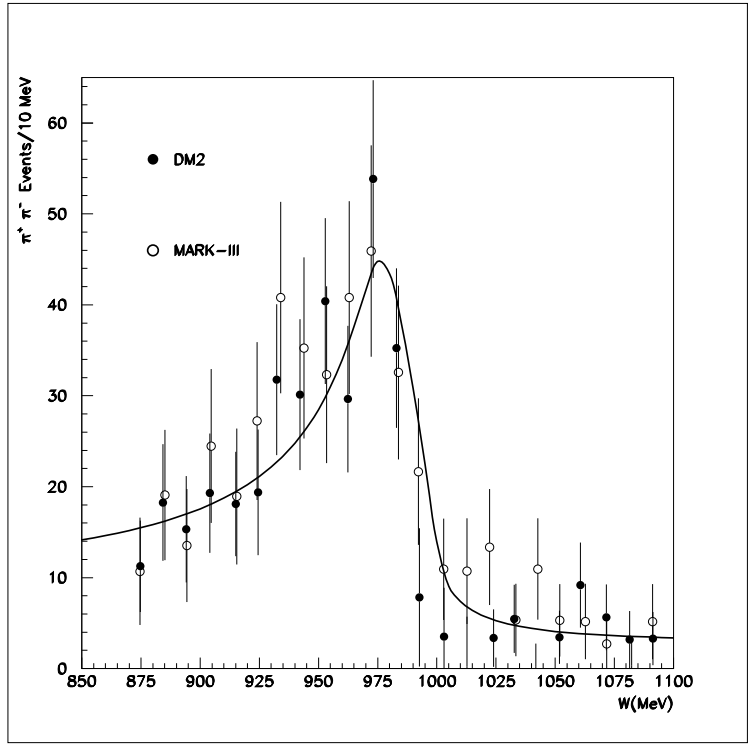}}
\vskip 0.6cm 

\caption[]{$\pi^+\pi^-$ event distribution in the $J/\Psi \rightarrow \phi \pi^+ \pi^-$
decay around the $f_0(980)$ mass. The
width of the bin is 10 MeV. }\label{fig:10}
\end{figure}

\begin{figure}[htb]
\centerline{
\epsfysize=3.9in 
\epsffile{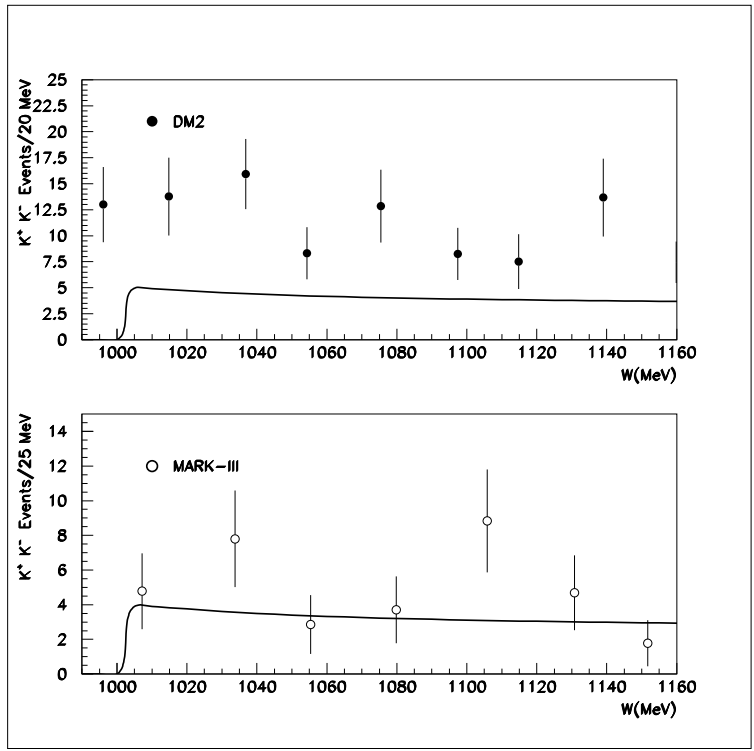}}
\vskip 0.6cm 

\caption[]{$K^+K^-$ event distribution in the $J/\Psi \rightarrow \phi K^+ K^-$ decay. 
The
upper panel corresponds to the data from DM2 \cite{dm2}, 20 MeV bins. The lower one
corresponds to the data from MARK-III \cite{mk3}, 25 MeV bins.}
\label{fig:kk}
\end{figure}

\medskip\noindent
While the data of DM2~\cite{dm2} have been published, this is not the case for the 
data of MARK-III~\cite{mk3}. On the other hand, both experiments, 
see Figs.\ref{fig:25} and \ref{fig:10}, are  compatible for the $\pi^+\pi^-$ 
event distribution
for 25 and 10 MeV bins\footnote{The data for the 10
MeV bins of MARK-III has been taken from Ref.\cite{pen}.}. However, this is not the
case for the $K\overline{K}$ event distribution, see Fig.\ref{fig:kk}. Consequently,
we will  only consider for
our fits the data from DM2~\cite{dm2} for the $\pi^+\pi^-$ event distributions in the 
$J/\Psi\rightarrow \phi \pi^+\pi^-$ decay, both for the 10 and the 25 MeV bins.
In fitting the data for the $J/\Psi \rightarrow \phi \pi^+\pi^-$ distribution from DM2 we
will fix $L_5^r$ and $L_8^r$ at the values given in Eq.(\ref{bij}). On the other hand,
$L_4^r$ and $L_6^r$ will be taken as free parameters. In this way, our expression for the
event distribution of the pions and kaons will have four free parameters:
${\mathcal{C}}_\phi$, $\lambda_\phi$, $L_4^r$ and $L_6^r$. However, there is still too much freedom
in fitting the data with this set of free parameters since the global normalization
constant ${\mathcal{C}}_\phi$ can only be determined within a large range and with
a sizeable
uncertainty. We will further restrict our fit by requiring that we can also describe the
pion event distribution in the $J/\Psi \rightarrow \omega \pi^+\pi^-$ decay, at least in 
the low energy region where the S-wave contribution, the one which we are
considering here, is  dominant. This extension of our model to the $\omega$
case is discussed in the next subsection. Imposing this requirement,
${\mathcal{C}}_\phi$ is fixed,\footnote{There is approximately a factor 1.11 between the global
normalization constant required for the DM2 data with respect the one
required for MARK-III.}   ${\mathcal{C}}_\phi=(16 \pm 3)\,$MeV$^{-1}$,
and the fitted values for
$\lambda_\phi$, $L_4^r (1.08~{\rm GeV})$ and $L_6^r (1.08~{\rm GeV})$ are:
\be
\label{result}
\lambda_\phi=0.17\pm 0.06~, \quad 10^3 \,L_4^r (1.08~{\rm GeV}) 
=0.44\pm 0.11~,\quad 10^3 \, L_6^r (1.08~{\rm GeV}) = -0.38\pm0.06~,
\ee
with a $\chi^2/{\rm dof} =0.92$. 
Clearly $\lambda_\phi\neq 0$, in contradiction with the OZI rule.
Furthermore, the pion event
distribution turns out to be very sensitive to the large $N_c$ subleading
low energy constants $L_4^r$ and $L_6^r$. The theoretical uncertainties given
in Eq.(\ref{result}) are obtained in the following way.
We have allowed for a relative change of  20\% in
the global normalization constant ${\cal C}_\phi$ when considering the 
data
with the $\phi$ and also the ones with 
the omega in the final state. We consider this estimate of the error in 
${\cal C}_\phi$  as conservative, since the ensuing  deviation from the $\omega \pi^+
\pi^-$ data for  such changes in  ${\cal C}_\phi$ is larger than the
uncertainty in the omega data by assuming a Poisson distribution. On the other hand, we 
have also allowed an uncertainty of $\pm 0.1$ GeV in the determination
of $q_{\rm max}$ from ref.\cite{npa} and then we calculated the band
of values for $\lambda_\phi$, $L_4^r$ and 
$L_6^r$. All these sources of uncertainty are added in
quadrature together with the statistical error 
given by the fitting procedure when using the central values for
${\cal C}_\phi$ and  $q_{\rm max}$.
 
\medskip\noindent
It is instructive to compare the values for the low energy constants
found here, e.g. Eqs.(\ref{result}), with the ones given in
Eq.(\ref{est}). While $L_6^r$ agrees perfectly within error bars, the sign of
$L_4^r$ is changed. Stated differently, if we evaluate from Eq. (\ref{result}) 
$10^3 \,L_4^r$ at the rho mass, we find a value of $0.71$, which is sizeably larger in 
magnitude than the central value given in Ref.\cite{daphne}. Also, it is larger than 
the positive value deduced from SU(2) information given
in Eq.(\ref{bij98}), at the scale $\mu=m_\rho$ one has for this case 
$10^3\,L_4^r=0.46\pm 0.25$, although both values are consistent within errors.
We reiterate that using large $N_c$ arguments, one would expect $L_4^r$ to be
zero (at a scale somewhere in the resonance region).  Therefore, our
increased value and also the one from Eq.(\ref{bij98}), clearly signals OZI violation. 
Quite 
differently, our value for $10^3 \,L_6^r (m_\rho) = -0.22$ is completely consistent with
the previous determination~\cite{daphne}. However, it does not fulfill
the positivity constraint $10^3 \,L_6^r (m_\rho) \ge 0.20$~\cite{stern,mouss} but it 
fulfills the other reasonable `positivity' bound previously discussed 
$10^3 \,L_6^r (m_\rho) \ge -0.59$. In fact, our value for $10^3\,L_6$ lies in an 
natural intermediate region between both extreme lower bound. Nevertheless, we have
also performed a series of fits enforcing the former constraint. We can
fit the $\phi$ data, but on the expense of very large values for
$\lambda_\phi$ and $L_4^r$. Furthermore, it is not possible to
simultaneously get a description of the $\omega$ decay data. We think that further study
 is needed in order to apply the positivity of the Dirac measure and also, we should 
stress that the LEC $L_6^r$ is plagued by the Kaplan--Manohar ambiguity~\cite{KM}.
It is important to point out
that one can criticize our determinations of the low energy constants
for two reasons. First, our model for the $J/\Psi$ decay with the
$\phi$ meson as a spectator is fairly simple, one could e.g. write down higher
order transition operators which would complicate the analysis. Given,
however, the fact that we can precisely reproduce the data both for the $\phi$ and the 
$\omega$ resonances, it is not 
obvious a priori that such a modified ansatz would lead to very
different results. Second, the use of unitarity to determine the scalar
form factors beyond one loop accuracy induces some inevitable model
dependence. To overcome this, one could think of doing a pure CHPT
analysis on the left wing of the scalar resonance. We believe,
however, that the present data in this energy region are not precise
enough for an accurate determination of the LECs. Independently of these
reservations, our analysis clearly underlines that the OZI rule is
strongly violated in the scalar $0^{++}$ sector, as indicated e.g. by
the large positive value of the LEC $L_4^r (m_\rho)$ and also by the non-vanishing 
value of $\lambda_\phi$. With respect the latter point, see also  the
footnote in section~\ref{sec:om}.

\medskip\noindent
It is also worth to indicate that in ref.\cite{paquito}, making use of the Inverse 
Amplitude Method (IAM) \cite{truong} with complete next-to-leading order CHPT strong 
amplitudes, a fit to the $I=0$ and $2$ S-wave and $I=1$ P-wave $\pi\pi$ and 
$K\overline{K}$ partial
 wave amplitudes was done in terms of the low energy coupling constants $L_1^r$, 
$L_2^r$, $L_3^r$, $L_4^r$, $L_5^4$ and $2 L_6^r+L_8^r$. This study has in common with 
the 
present one that a complete matching to the relevant next-to-leading order CHPT results 
was given and at the same time fully unitarity amplitudes were derived. The 
experimental data was very well reproduced up to energies around 1.2 GeV giving rise to 
the presence of the resonances $\rho$ and $f_0(980)$. The values obtained for $L_5^r$ 
and $2 L_6^r+L_8^r$, within errors, are consistent with those recently obtained in ref. 
\cite{bij}. This implies agreement of the results of that reference with our choice 
for the values of  $L_5^r$ and $L_8^r$ \cite{bij} and our presently
determined value for $L_6^r$.
The main difference between the set of values given in ref.\cite{paquito} and those in 
ref.\cite{bij} corresponds to the value of $L_2^r$. While in the former case 
$L_2^r\simeq 2\,L_1^r$ as required by Vector Meson Dominance (VMD)
\footnote{There is a very close link 
between VMD and the IAM in the vector channels \cite{truong2,nd}.}\cite{reso}, this 
relation is only fulfilled within errors by the values given in ref.\cite{bij}. 
Nevertheless, in ref.\cite{paquito} $10^3\, L_4^r(m_\rho)=0.2\pm 0.1$. This 
value, although on the positive side, is  
incompatible with our present one, $10^3\,L_4^r(m_\rho)=0.71\pm 0.11$, and, 
within errors, is compatible with the rest of analyses presented in this section except 
for~\cite{mouss2}. Summarizing, for
$L_6^r$ there is a rather good agreement between our present study and
refs.\cite{gl1}, \cite{bij}, \cite{paquito} in disagreement with the finding of ref. 
\cite{mouss,mouss2}. On the other hand, for $L_4^r$ our present analysis is compatible 
only with that value of $L_4^r$ determined from ${\mathcal{O}}(p^6)$ CHPT 
\cite{bij98,bij} which 
also find quite  a sizeable central value at $\mu=M_\rho$, around $0.5\times 10^{-3}$, different 
from the smaller numbers of refs.\cite{gl1,mouss,mouss2,paquito}.

\begin{figure}[htb]
\begin{center}
\epsfysize=3.8in 
\epsffile{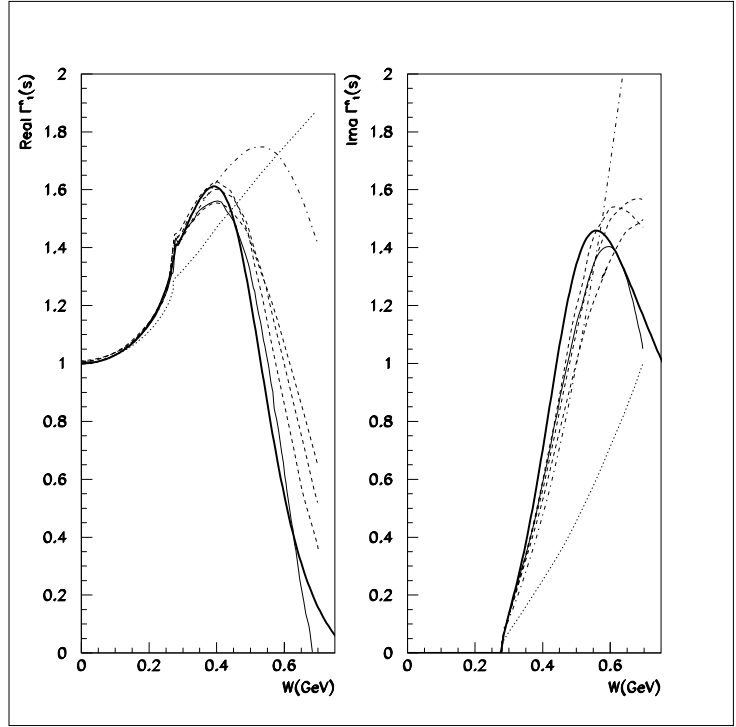}
\end{center}
\vskip 0.7cm 

\caption[]{Normalized non--strange scalar form factor of the pion
  (wide solid line). Dotted and dot--dashed lines: One and two loop CHPT
  results, respectively. The three dashed lines are the dispersion
  theoretical results from Ref.\cite{dgl}. The thin solid lines represent the 
exponentiated two--loop CHPT results~\cite{gm}. Our results correspond to the thick solid
 lines. Left/right panel: Real/imaginary part. 
}\label{fig:sff1}
\end{figure}

\begin{figure}[htb]
\centerline{
\epsfysize=3.in 
\epsffile{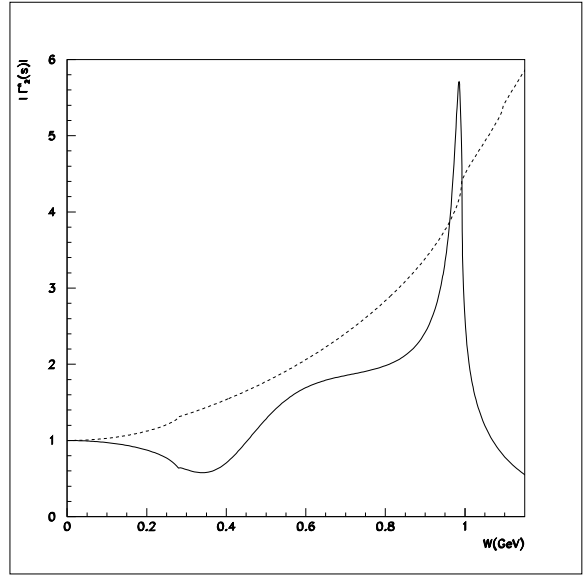}}
\vskip 0.7cm 
\caption[]{Normalized non--strange scalar form factor of the kaon
(solid line) in comparison to the one loop CHPT result (dashed line).
}\label{fig:sff2}
\end{figure}

\begin{figure}[htb]
\centerline{
\epsfysize=3.8in 
\epsffile{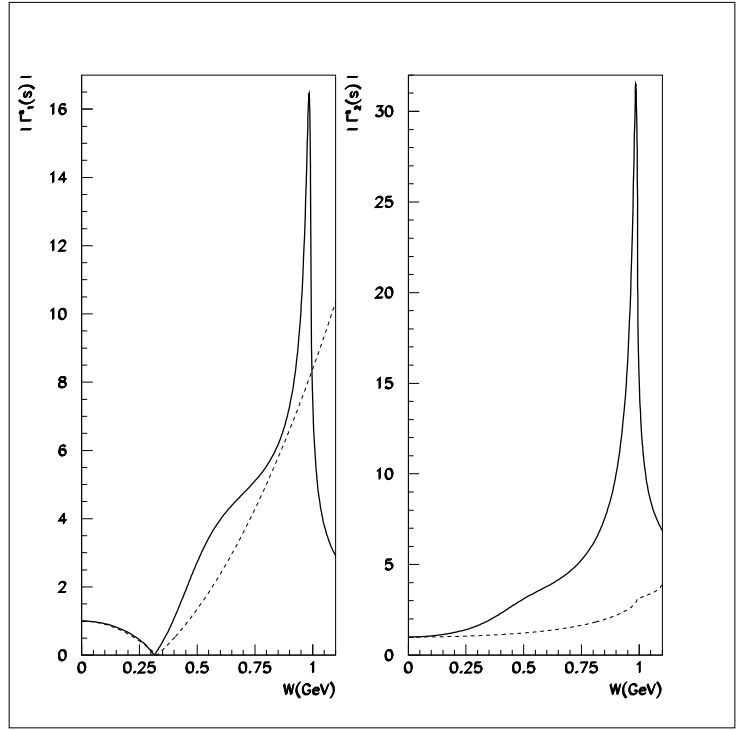}}
\vskip 0.7cm 
\caption[]{Normalized strange scalar form factor of the pion (left
panel) and  of the kaon (right panel). Solid lines: Chiral unitary
approach. Dashed lines:  one loop CHPT result.
}\label{fig:sff3}
\end{figure}

\medskip\noindent
The resulting non--strange and strange normalized scalar form factors
of the pion and the kaon are shown in Fig.\ref{fig:sff1},  Fig.\ref{fig:sff2} and
Fig.\ref{fig:sff3}. In the case of the non--strange scalar form factor,
we show for comparison in Fig.\ref{fig:sff1} the one-- and two--loop CHPT~\cite{gl1,gm} 
as well as the dispersion theoretical results~\cite{dgl} and the exponentiated 
two--loop CHPT result~\cite{gm}. In the latter case, the two--loop CHPT result is 
improved by making use of an Omn\`es resummation in terms of the next-to-leading order 
CHPT phase shifts. Our result is close to the
ones obtained by a different method in Ref.\cite{dgl} and even closer to the 
exponentiated two--loop CHPT results of ref. \cite{gm}. The agreement is worse when 
comparing
 our results with the so called modified-Omn\`es representation of refs. \cite{gm,bij98}.
 We also remark that the
two--loop representation covers the main feature of this quantity below $W
\simeq 600$~MeV, as it is known since long~\cite{gm}. The strange
scalar form factor of the pion is reasonably well described for energies below
350~MeV. In contrast, the strange and the non--strange scalar form factor of
the kaon are poorly described at one loop, as expected from the larger
mass of the kaon.

\medskip\noindent
In Figs.\ref{fig:25} and \ref{fig:10},
 we show the curves from ours fit to the $\pi^+\pi^-$ event distribution 
in comparison with the experimental data from DM2 and MARK-III  for the 25 and 
the 10 MeV bins, respectively. The data of
MARK-III have been multiplied by  the factor
${\mathcal{C}}^2_{\phi, \rm DM2}/{\mathcal{C}}^2_{\phi, \rm MARK-III}\simeq 1.11^2$ 
in order to facilitate the comparison 
between both sets of data. The agreement with the experimental data is 
very good as indicated by the low $\chi^2$/dof of 0.92. By comparing
Fig.\ref{fig:25} with the left panel of Fig.\ref{fig:sff3} we see that the
event distribution of the two pions is dominated by the strange scalar
form factor of the pion. Moreover, in Fig.\ref{fig:25} the changes in 
the results when allowing a change in the cut-off $q_{\rm max}$ by $\pm 0.1$ GeV \cite{npa} 
are shown to be quite small.  In 
Fig.\ref{fig:kk} our prediction for
the $K^+K^-$ event distribution is depicted. Incidentally, we find better agreement 
with the data of 
MARK-III than with the ones of DM2 with respect to this decay mode.
This was also noted in Refs.\cite{pen,Juel}. In fact, in these references a fit of 
similar quality to the data of DM2 and MARK-III is also given. The important difference 
between 
their method and ours is that we have devised a dynamical approach which means that 
the parameters that enters in our description of the problem are not specific to it and 
can be related to many other physical observables. This is particularly true for $L_4^r$
 and $L_6^r$. But even for $\lambda_\phi$ we will see in the next subsection how it can 
be related to the whole set of U(3) processes that follows from the decays of the 
$J/\Psi$ resonance to any vector 
resonance belonging to the lightest nonet of vector resonances and two pseudoscalars. On
 the other hand, in ref. \cite{pen,Juel} no attempt was done to describe the 
$J/\Psi \rightarrow \omega \pi\pi$ decays.

\subsection{The $\omega$-meson case}
\label{sec:om}

A priori one can expect that the $J/\Psi \rightarrow \omega \pi\pi$ decay 
requires a rather 
different dynamical description than that for the mode
$J/\Psi \rightarrow \phi \pi\pi$ considered so far. 
For instance, the approach of considering the $\omega$ as
a spectator is by no means so clear as for the $\phi$ case. 
Note that the Dalitz plot for this decay has  very
clear bands due to the decays $b_1 (1235) \to \omega \pi$ and $f_2
(1270) \to \pi^+\pi^-$~\cite{mk3}. The latter induces a
sizeable D-wave contribution so that our approach can only be applied
for the first few hundred MeV of the two pion event distribution.
However, making use of SU(3) symmetry, we can  extend our considerations
from section~\ref{sec:model}
and we will present our calculated S-wave contribution to the invariant mass 
distribution of the pions in the $J/\Psi \rightarrow \omega
\pi\pi$ decay. This calculation is completely fixed in terms of the parameters already
given for the $\phi$ case, cf. Eq.(\ref{result}), except for the
global normalization constant for which we have also used the experimental data from 
the $J/\Psi \rightarrow \omega \pi^+\pi^-$ decays to further constraint its value.

\begin{figure}[htb]
\centerline{
\epsfysize=3.8in 
\epsffile{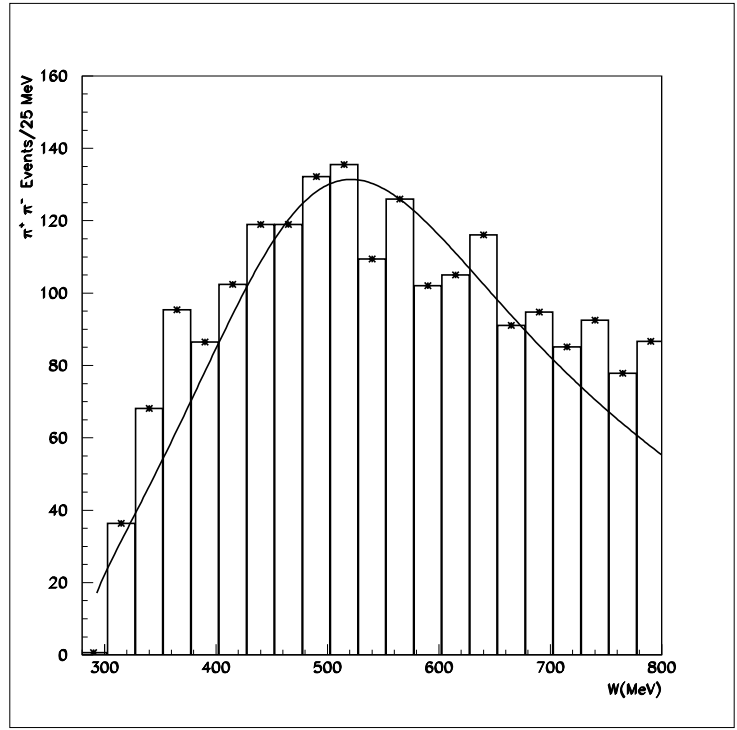}}
\vskip 0.4cm 

\caption[]{$\pi^+\pi^-$ event distribution in the $J/\Psi \rightarrow \omega \pi^+ \pi^-$ 
decay. The width of the bin is 25 MeV. Only the S-wave contribution is calculated.
The onset of the D-wave contribution can be seen at energies larger than 700 MeV.
}\label{fig:w}
\end{figure}

\medskip
\noindent
Let us now be more specific and work out the aforementioned relation between
the two cases.  An invariant SU(3) Lagrangian involving the octet and
singlet of vector resonances 
$V^{(8)}_\mu$ and $V^{(1)}_\mu$, respectively, and the corresponding ones of the scalar 
sources $S^{(8)}$ and $S^{(1)}$ can be written as (making also use of Lorentz
invariance) 
\be
\label{lag}
{\mathcal{L}}=\hat{g} \, \biggl(\Psi^\sigma \langle V^{(8)}_\sigma S^{(8)}\rangle
+\nu \Psi^\sigma V_\sigma^{(1)} S^{(1)}\biggr)~,
\ee
where $\hat{g}$ is an overall coupling constant whose precise value is not needed in
the following. We only need to determine the relative strength of the 
octet to singlet couplings given in terms of the real parameter $\nu$. The
symbol $\langle...\rangle$ refers to the trace over the SU(3) indices of 
the matrices $V^{(8)}$ and  $S^{(8)}$. These are defined via
\be
\label{vmu}
V_\sigma^{(8)} = \left[ \matrix{\frac{1}{\sqrt{2}}\rho^0+\frac{1}{\sqrt{6}}V_8 & 
\rho^+ & K^{*\,+} \cr \rho^- & -\frac{1}{\sqrt{2}}\rho^0+\frac{1}{\sqrt{6}}V_8 & 
K^{*\,0} \cr K^{*\,-} & \bar K^{*\,0} & -\frac{2}{\sqrt{6}}V_8} \right]_\sigma~.
\ee
and similarly for the $S^{(8)}$ matrix. In the previous equation we have denoted by $V_8$
the $I=0$ state of the octet of vector resonances. This formalism is  in close 
analogy with the one used in CHPT for the octet of pseudoscalars, compare Eq.(\ref{Phi}). 
Denoting by $S_8$ the $I=0$ operator of the octet of scalar sources, we can write the 
terms involving $V_8$ and $V_1$ of Eq.(\ref{lag}) as:
\be
\label{terms}
\Psi^\sigma \left(V_{8\,;\sigma} S_8+\nu V^{(1)}_\sigma S^{(1)}  \right)~.
\ee
Considering ideal mixing\footnote{In Ref.\cite{okonor} the departure from the ideal
mixing in the $\omega-\phi$ system is thoroughly studied comparing different models, and
is described by a rotation of the ideal mixing states
with a rotation angle $|\delta_V|\approx 3^0$. This departure would produce corrections 
of the order of a 5$\%$ with respect the ideal mixing situation considered here. However,
this value should be compared with the departure from 1 of the parameter $\nu$ which from
Eq. (\ref{lphi}), taking into account the value of $\lambda_\phi$ given in Eq. 
(\ref{result}), is about $40\%$, a much bigger effect than any expected deviation from the ideal
mixing situation in the $\omega-\phi$ system. Note that $\nu=1$ is the value expected 
from U(3) symmetry and for this case $\lambda_\phi=0$ and $\lambda_\omega=\infty$, see Eqs. 
(\ref{lphi},\ref{lw})} between the $V_8$ and the $V^{(1)}$ then:
\be
\label{im}
V_8=\frac{\omega}{\sqrt{3}}-\sqrt{\frac{2}{3}}\phi~, \quad
V^{(1)}=\sqrt{\frac{2}{3}}\omega+\frac{\phi}{\sqrt{3}}~.
\ee
In an analogous way we will also introduce the scalar sources $S_\omega$ and $S_\phi$
defined by
\be
\label{ssdd}
S_8=\frac{S_\omega}{\sqrt{3}}-\sqrt{\frac{2}{3}}S_\phi~, \quad
S^{(1)}=\sqrt{\frac{2}{3}}S_\omega+\frac{S_\phi}{\sqrt{3}}~.
\ee
Note that in a quark model language, consistently with the transformation properties under
SU(3), we can write:
\be
\label{ssq}
S_\phi=\bar{s}s\;\;\;\;\hbox{and}\;\;\;\; S_\omega=\frac{1}{\sqrt{2}}(\bar{u}u+\bar{d}d)
~.
\ee
Rewriting Eq.(\ref{terms}) in terms of $\omega$, $\phi$, $S_\omega$ and $S_\phi$, we
have:
\be
\label{su3f}
\frac{2+\nu}{3}\Psi^\sigma \phi_\sigma \left(S_\phi+S_\omega 
\frac{\sqrt{2}(\nu-1)}{2+\nu} \right)
+ \frac{\sqrt{2}(\nu-1)}{3}\Psi^\sigma \omega_\sigma \left(S_\phi+S_\omega 
\frac{1+2\nu}{\sqrt{2}(\nu-1)} \right) ~.
\ee
In this way, the parameter $\lambda_\phi$ introduced in Eqs.(\ref{nquarks},\ref{S}) and 
fitted in the previous subsection, can now be expressed as:
\be
\label{lphi}
\lambda_\phi=\frac{\sqrt{2}(\nu-1)}{2+\nu}~.
\ee
>From this equation we can isolate $\nu=\nu(\lambda_\phi)$ and then predict the
corresponding parameter $\lambda_\omega$,
\be
\label{lw}
\lambda_\omega=\frac{1+2\nu ( \lambda_\phi ) }{\sqrt{2}(\nu ( \lambda_\phi )-1)}
\ee
We can also obtain from Eq.(\ref{su3f}) the global normalization constant for the 
$\omega$ in terms of the one of the $\phi$ since
\be
\label{rat}
\frac{C_\omega}{C_\phi}=\frac{\sqrt{2}(\nu-1)}{2+\nu}=\lambda_\phi~.
\ee 
In Fig.\ref{fig:w} we show the calculated S-wave contribution to the event distribution of 
pions in the $\Psi \rightarrow \omega \pi^+\pi^-$ decay. This calculation does not
introduce any new free parameter since all of them have been fixed in terms of the one of
the $\phi$ case by Eqs.(\ref{lw}) and (\ref{rat}). The description of the data of
MARK-III up to around 0.7 GeV is very good. For higher energies the D-wave contribution 
cannot be further neglected.

\section{Conclusions}
\label{sec:concl}

In this work we have addressed the problem of the $J/\Psi$ decays into a
vector ($\phi, \omega$) and two
pseudoscalar mesons (Goldstone bosons)
measured at DM2~\cite{dm2} and MARK-III~\cite{mk3}. These processes are
considered to be mediated by the corresponding scalar form factors
of the pseudoscalar mesons if one considers the emitted vector meson as a
spectator. Consequently, these
reactions are rather interesting since they are very sensitive to OZI violation
physics, in our scheme parameterized by the constants $\lambda_\phi$, see Eq.(\ref{S}),
and the low energy constants $L_4^r$ and $L_6^r$ of chiral
perturbation theory.
The first of these constants parameterizes the direct admixture of non--strange
quarks to the scalar interpolating field for our model of the $J/\Psi$ decay
with the $\phi$ playing the role of a spectator, see
Figs.\ref{fig:OZI},\ref{fig:anat}.
The two low energy constants enter the one loop description of the
pion and kaon scalar form factors. To describe these properly for the
range of energies relevant here, we have combined
information coming from next-to-leading order (one loop) chiral perturbation
theory (CHPT) with the unitarity 
requirements which are valid to all orders in the chiral expansion. In addition, we also have
calculated for the first time the next-to-leading order CHPT
kaon scalar form factors, for strange, $\bar{s}s$, and non--strange,
$\bar{u}u+\bar{d}d$, scalar--isoscalar quark densities.
The unitarity requirements were imposed by using the strong $I=0$ $\pi\pi$ and $K\overline{K}$ 
amplitudes derived in Ref.\cite{npa}. The amplitudes given in that paper not 
only describe accurately the S-wave $I=0$ and $I=1$ strong scattering data but also have 
been used  to successfully reproduce or even predict 
experimental data for the whole set of reactions listed in the
Introduction. With this input, we have successfully described, from
threshold up to around 1.2 GeV,\footnote{In principle, one could also go to
higher energies but that is much more difficult due a) to the appearance
of multiparticle states and b) due to novel interaction vertices as
discussed in section~\ref{sec:model}.} the event
distribution of the $\pi^+\pi^-$ system in the $J/\Psi \rightarrow \phi \pi^+\pi^-$ decay. 
We have then predicted, in agreement with the data from MARK-III, the event distribution of 
kaons in the $J/\Psi \rightarrow \phi K^+ K^-$ reaction and the low energy part, where 
the S-wave dominates, of the event distribution of $\pi^+\pi^-$ pairs in  the
$J/\Psi \rightarrow \omega \pi^+\pi^-$ decay. Furthermore, the OZI violation parameter
$\lambda_\phi$ comes out different from zero. This also holds for the
low energy constants $L_4^r$ and $L_6^r$. While the value of the
latter agrees with previous estimates \cite{gl1,paquito}, our result for $L_4^r$ is
sizeably larger in magnitude as most previous estimations 
\cite{gl1,mouss,mouss2,paquito}. However, it is compatible 
within errors  with the quite constraint value derived by combining the information 
from ${\mathcal{O}}(p^6)$ 
SU(2) \cite{bij98} and SU(3) \cite{bij} CHPT. This offers
another indication that the OZI rule does not account for the physics
in the scalar $0^{++}$ channel, as stressed e.g. in Refs.\cite{isgur,stern,isgth}.
The scheme employed here offers  a unique approach  to describe  the
scalar sector, which has been  at the heart of many investigations over the
last decade.

\medskip\noindent
Clearly, to further improve the approach presented here, it would be mandatory
to not only have event distributions but rather normalized data. This would allow 
one to pin down the low energy constants $L_4^r$ and $L_6^r$ more precisely together 
with  the OZI violation parameter $\lambda_\phi$ as well as the product 
$(\hat{g} \, m_q \, B_0)^2$ (i.e. normalization of the scalar form
factors times the strength of the scalar source to vector meson coupling). 
With respect to the former, one could also try to do a pure 
one (or even two) loop CHPT calculation for small two--pion invariant masses, i.e. on the
left wing of the $f_0 (980)$ resonance. Clearly, the presently available data are not
precise enough for successfully doing that, but such a program is in
principle of the similar interest as the study of chiral dynamics in $\tau$ decays, see
e.g. Refs.\cite{CU,stern2}, specially when referring to the scalar sector.

\subsection*{Acknowledgments}
J.A.O. would like to acknowledge stimulating discussions
with T. Barnes. U.-G.M. is grateful to V. Bernard for some
pertinent comments. The work of J.A.O. was supported in part by funds from
DGICYT under contract PB96-0753 and from the EU TMR network Eurodaphne, contract no. 
ERBFMRX-CT98-0169.

\bigskip
\newpage

\newpage

\section*{Figures}

\vskip 1cm

\vfill

\end{document}